\documentclass[12pt]{iopart}
\usepackage{epsf}
\begin{document}

\title{The Quantum Self-eraser}

\author{Jes\'us Mart\'{\i}nez-Linares\dag\
\footnote[3]{(jesusmartinezlinares@hotmail.com)} and Julio Vargas
Medina\ddag  }

\address{\dag\ Facultad de Ingenier\'{i}a en Tecnolog\'{i}a de la Madera. \\
P.O. Box 580, Universidad Michoacana de San Nicol\'{a}s de
Hidalgo. 58000 Morelia, Michoac\'{a}n, M\'{e}xico.}

\address{\ddag\ Centro de Investigaci\'{o}n y Desarrollo del Estado de Michoac\'an \\
Avenida Ju\'{a}rez s/n. 58060 Morelia, Michoac\'{a}n, M\'{e}xico}

\begin{abstract}
A scheme for an atomic beam quantum self-eraser is presented. The
proposal is based on time reversal invariance on a quantum optical
Ramsey fringes experiment, where a realization of complementarity
for atomic coherence can be achieved. It consists of two high
finesse resonators that are pumped and probed by the same atom.
This property relates quantum erasing with time reversal symmetry,
allowing for a full quantum erasing of the which-way information
stored in the cavity fields. The outlined scheme also prepares and
observes a non-local state in the fields of the resonators: a
coherent superposition between correlated states of
macroscopically separated quantum systems. The proposed scheme
emphasizes the role of entanglement swapping in delayed-choice
experiments. Finally, we show that the quantum self-eraser
violates temporal Bell inequalities and analyze the relation
between this violation and the erasability of which-way
information.
\end{abstract}

\pacs{42.50.-p  03.67.-a 42.50.Xa} \hspace{2.5cm}{\it Keywords:}
{\small Complementarity, quantum eraser, cavity-QED, Bell ineq.}


\section{Introduction}

The complementarity principle has traditionally been addressed in
order to illustrate the role of observer in Quantum Mechanics. One
of the most intriguing aspect of the issue is the possibility of
delayed-choice experiments \cite{[1]}. According to it, the
experimenter may decide for instance to display wavelike aspects
of an atomic system, long after it has been forced into a
particle-like behavior \cite{[2]}. If the atomic state is
entangled with a memory system available to the observer, i.e., a
which-way detector (WWD), the experimenter may decide to postpone
his decision even after the atom has been detected. Interference
fringes can be recovered, provided the which-way information (WWI)
is erased from the WWD \cite{Englert00}. An experimental
demonstration of a quantum eraser have been given using entangled
photon pairs \cite{Herzog95,Scully00,Bjork00}, using single photon
entanglement between spatial mode and polarization state
\cite{Englert99_1}, and using atom interferometry \cite{Durr98}.

The great progress in cavity quantum electrodynamics offers the
possibility of designing a quantum detector to retain which-way
information for rather long time after the system has been
detected \cite{Walther94}. On the other hand, microwave resonators
can couple strongly to Rydberg atoms. As a matter of fact,
high-finesse microwave resonators have been used to generate
Einstein-Podolsky-Rosen pairs of atoms \cite{Hagley97} and in the
construction of quantum logic-gates \cite {Turchette95}. Thus,
quantum optical Ramsey interferometers are good candidates for
complementarity experiments. Several models of cavity-QED methods
for demonstrating complementarity have been proposed. Some of them
supplement the Young double-slit experiment with WWD based on
dispersive atom-field interaction
\cite{Haroche92,Storey93,Gerry96}. Other proposal uses two atomic
beams as WWD of each other using a cavity field in order to
entangle them \cite{Bogar96}. Dispersive cavity-QED experiments
were reported in \cite{Brune96} and more recently in
\cite{Bertet01}, where the quantum-classical limit of
complementarity have been explored using a cavity field
continuously tuned from microscopic to macroscopic regimes. A
simplified model for quantum erasing have been found in
\cite{Zheng00}.

We present in this paper a novel quantum-erasing scheme:\ the
quantum self-eraser, based on a different strategy. The innovative
idea behind is to use time reversibility to ''undo'' the
interactions that deposited the WWI into the WWD. Thus, we launch
back the same atom (or a velocity reversal replica of it) to act
as the eraser atom (Erason). Controlling time reversibility, we
can let the Erason absorb or erase its own WWI that was deposited
previously on the WWDs. This scheme will emphasize the role of
entanglement swapping in quantum erasing operation. Moreover, the
phase difference $\phi $ between the interfering alternatives is
mapped into the non-local phase of a superposition between the two
macroscopically separated cavity-field states. The self-eraser
probes this phase, allowing for a joint demonstration of
complementarity and the generation of a non-local coherent
macroscopic superposition. Finally, we analyze the relation
between the violation of temporal Bell inequalities as expression
of the establishment of macroscopic coherence, and the erasability
of the WWI in the quantum self-eraser.

The paper is organized as follows. In section 2, complementarity
in a quantum optical Ramsey interferometers (QORI) is discussed in
order to set up the notation. In section 3, we prepare the stage
for the discussion showing how the QORI scheme can be used to
prepare in an unitary fashion a non-local field state. Section 4
is devoted to probe of the phase of the macroscopic superposition
via quantum self-erasing. Section 5 studies the violation of
temporal Bell inequalities. Finally, we end up with a discussion
and a summary of our results.

\section{The quantum optical Ramsey interferometer}
In this section, we describe a quantum optical Ramsey
interferometer (QORI). We can summarize the system as follows. It
consists in an atomic interferometer, as depicted in Figure 1a.
The interfering ways are realized on the internal states of a
Rydberg two-level atom, that we denote by $\left\{ |a\rangle
,|b\rangle \right\} $. Before entering into the interferometer,
the atoms are prepared in the upper state $|a\rangle .$ A
microwave resonator provides the quantized beam splitter (BS) as
well as the WWD. Following the strategy of \cite{[8]}, the phase
shifter (PS) is provided by an electrostatic external field
applied in the region between the cavities. The differential Stark
shift between upper and lower levels induces a relative phase
$\phi $ that enables an external control of the off-diagonal
elements of the density matrix. Subsequently, a classical
microwave -$\frac{\pi }{2}$ pulse provides the beam merger
(BM).The combined interferometer action performs a transformation
from the initial state

\begin{equation}
\rho ^{o}=\frac{1}{2}\left( 1+\sigma _{z}\right) \otimes \rho
_{D}^{o}. \label{ initial state QD}
\end{equation}
to the final state
\begin{eqnarray}
\rho ^{f} &=&\frac{(1-\sigma _{x})}{2}\;C\rho _{D}^{o}C+\frac{(1+\sigma _{x})%
}{2}\;a_{{}}^{\dag }S\;\rho _{D}^{o}\;Sa  \label{final state QD} \\
&+&\frac{(\sigma _{z}+i\sigma _{y})}{2}\;C\;\rho _{D}^{o}\;Sa\;e^{-i\phi }+%
\frac{(\sigma _{z}-i\sigma _{y})}{2}\;a_{{}}^{\dag }S\;\rho
_{D}^{o}C\;e^{i\phi }\,,  \nonumber
\end{eqnarray}
where the operators \cite{[8]}
\begin{eqnarray}
S &=&\frac{\sin \left( \Omega \tau \sqrt{aa^{\dag }}\right)
}{\sqrt{aa^{\dag
}}}=S^{\dag },  \label{CSoperators} \\
C &=&\cos \left( \Omega \tau \sqrt{aa^{\dag }}\right) =C^{\dag },
\nonumber
\end{eqnarray}
have been introduced, $\left\{ \sigma _{x},\sigma _{y},\sigma
_{z}\right\} $ are the Pauli spin matrices of the two-level atom,
and $\rho _{D}^{o}$ is the initial state of the WWD, i.e., the
cavity-field inside the resonator. Here $a$ and $a^{\dagger }$are
annihilation and creation operators of the cavity field mode,
$\Omega $ is the vacuum Rabi frequency of the atom-field
interaction, and $\tau $ the interaction time, i.e., the time of
flight of the atom through the cavity. In the derivation of
(\ref{final state QD}) resonant interaction has been assumed, and
cavity damping and spontaneous atomic decay have been neglected.

\begin{figure}
\begin{center}
\epsfxsize=34pc 
\epsfbox{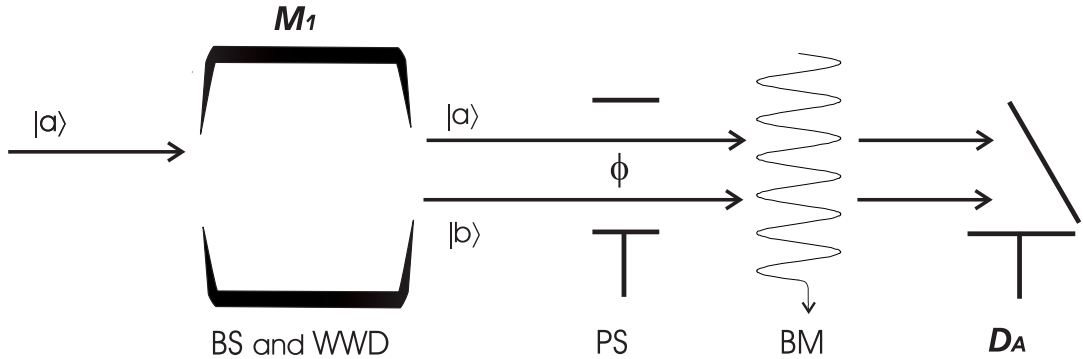}
\end{center}
\caption{\label{fig1}Quantum optical Ramsey interferometer. The
interfering ways are realized on the internal states $\left\{ \mid
a\rangle,\mid b\rangle\right\} $ of the Quanton: a two-level
Rydberg atom. A quantized BS provides also the WWD in cavity
M$_{1}$. After the phase shifter PS a classical microwave field
recombines the ways (BM). The state of the atom is measured
finally at $D_A$}
\end{figure}

The final state of the WWD is obtained after tracing out the
atomic degree of freedom. Thus
\begin{equation}
\rho _{D}^{f}=\rm{tr}_{atom}\left\{ \rho ^{f}\right\} =w_{+}\rho
_{D}^{+}+w_{-}\rho _{D}^{-},  \label{rhoDfinal}
\end{equation}
where $\rho _{D}^{+}$, $\rho _{D}^{-}$ are the detector's two
final states~ corresponding to each way. The WWD can be regarded
as a memory system, with two internal pointer states correlated to
the interfering ways, given by $\rho _{D}^{+},\,\rho _{D}^{-}$ in
(\ref {rhoDfinal}). The quantities
\begin{equation}
w_{\pm }\equiv \rm{tr}\left\{ \frac{1\mp \sigma _{x}}{2}\rho
^{f}\right\} , \label{w+-}
\end{equation}
are the probabilities for taking each way after the BS. Thus, the
asymmetry of the BS is measured by the predictability
\begin{equation}
\mathcal{P}=|w_{+}-w_{-}|.  \label{P}
\end{equation}

The final state of the exit atom is measured by means of state
selective field ionization techniques in $D_{A}$. Quantum optical
Ramsey fringes are exhibited in the probability of detecting the
outgoing atoms in a definite
internal state, in case complementarity allows for them. For instance, $%
\mathcal{P}_{aa}(\mathcal{P}_{ab}),$ the probability of detecting
the Quanton in the upper (lower) state can be calculated to yield
\begin{eqnarray}
\mathcal{P}_{aa} &=&\frac{1}{2}\left( \left\langle
C^{2}\right\rangle _{o}+\left\langle S^{2}aa_{{}}^{\dag
}\right\rangle _{o}+2\rm{Re}\left\{
\mathcal{C}e^{-i\phi }\right\} \right) ,  \label{Paa} \\
\mathcal{P}_{ab} &=&1-\mathcal{P}_{aa},  \nonumber
\end{eqnarray}
where the contrast factor $\mathcal{C}$ is given by the expression

\begin{equation}
\mathcal{C}=2\left\langle SaC\right\rangle _{o}.  \label{C}
\end{equation}
The averages in this equation are taken over the initial field
prepared inside the cavity, and can be calculated after an
appropriated expansion of the state in the photon number bases.

The stage is now ready for the analysis of complementarity in the
interferometer. Complementarity demands that distinguishability of
the ways must be followed by lost of coherence in the visibility
$\mathcal{V}$ of the
interference pattern in (\ref{Paa}). A measure of the distinguishability $%
\mathcal{D}$ of the ways in a two-ways interferometer have been
given in \cite{Englert96,Glauber79}, quantifying the maximum potential WWI
that can be available to the experimenter. Two sources of WWI\
contribute to $\mathcal{D} $. One is the a-priori WWI the
experimenter has about the ways, stemming from the asymmetric
preparation of the beam splitter, and thus, quantified by the
predictability $\mathcal{P}$ in (\ref{P}). The other source of WWI
depends on the quantum ''Quality'' $\mathcal{Q}$ of the WWD, i.e.,
its ability to trace down the ways of the two-level system
(Quanton) via quantum correlations. Recently, a formalism has been
developed \cite{Jesus01} which allows one to separate both
contributions, even in the case where beam splitter (BS) and WWD
are provided by the same physical interaction \cite{Rinton}.
Assuming pure state preparation in (\ref{ initial state QD}), we
have

\begin{equation}
\mathcal{D}=\sqrt{\left( 1-\mathcal{P}^{2}\right) \mathcal{Q}^{2}+\mathcal{P}%
^{2}}.  \label{Dedo}
\end{equation}

As can be seen from the above equation, a high-quality
$\mathcal{Q}=1$ WWD
implies full distinguishability $\mathcal{D}=1,$ no matter the value of $%
\mathcal{P}$ and viceversa. In particular, a perfect-quality WWD
will prevent any fringes from been displayed at the output port of
the interferometer, for any value of the predictability
$\mathcal{P}$. In fact, for pure state preparation the following
equality holds \cite{Rinton}
\begin{equation}
\left( 1-\mathcal{P}^{2}\right) \mathcal{Q}^{2}+\mathcal{P}^{2}+\mathcal{V}%
^{2}=1.  \label{1.1}
\end{equation}

The visibility can be rapidly calculated from (\ref{Paa}) to be
$\mathcal{V}^{{}}=\left| \mathcal{C}\right|$ \cite{Englert96},
where the contrast factor $\mathcal{C}$ ranging from 0 to 1 is
given in (\ref{C}). Fringes are degraded by the potential
availability of WWI stemming from the introduction of a quantum
WWD into the interferometer. Thus, the observation of an
interference pattern in the detection probability of the atom
depends on the state preparation of the cavities. If they are
unable to acquire which-way information ($\mathcal{Q}=0$), quantum
interference is observable due to the indistinguishability of the
path leading to the same final state. On the other hand, if we
prepare the cavities to act as which-way detectors, for instance
in a Fock state ($\mathcal{Q}=1$), then no interference effects
are observable ($\mathcal{V=}0$), as ensued by complementarity
\cite{Rinton}. As a matter of fact, $\mathcal{V}$ has been
measured in a recent experiment \cite{Bertet01} for different
preparation of the initial cavity field state, showing the
transition between both complementary situations.

Consider now the initial cavity field prepared in the vacuum state
$\rho _{D}^{o}=\mid 0\rangle \langle 0\mid $. The relevant
quantities of this section can be rapidly calculated to yield
\begin{eqnarray}
\mathcal{P}_{ab} &=&1/2,  \label{tochillo} \\
\mathcal{P}_{{}} &=&\left| \cos (2\Omega \tau )\right| ,  \nonumber \\
\mathcal{Q} &=&1\Rightarrow \mathcal{D=}1,\mathcal{V}=0.
\nonumber
\end{eqnarray}

No interference pattern can be measured at the output port of the
interferometer, since the cavity-field stores full WWI about the
alternatives taken by the atom ($\mathcal{Q} =1$). The upper
(lower) way is perfectly correlated to the no-photon (one photon)
state of the cavity field. In order to restore the interference
pattern, the WWI must be erased so it is no longer available to
the experimenter. This can be done by launching a second Rydberg
atom through the cavity, in order to absorb the "which-way" photon
stored in the cavity field. An interference pattern can be builded
in the correlated detection probability of both Quanton and Erason
atoms \cite{Englert00}.

\section{Unitary preparation of a non-local field state}

\begin{figure}
\begin{center}
\epsfxsize=32pc 
\epsfbox{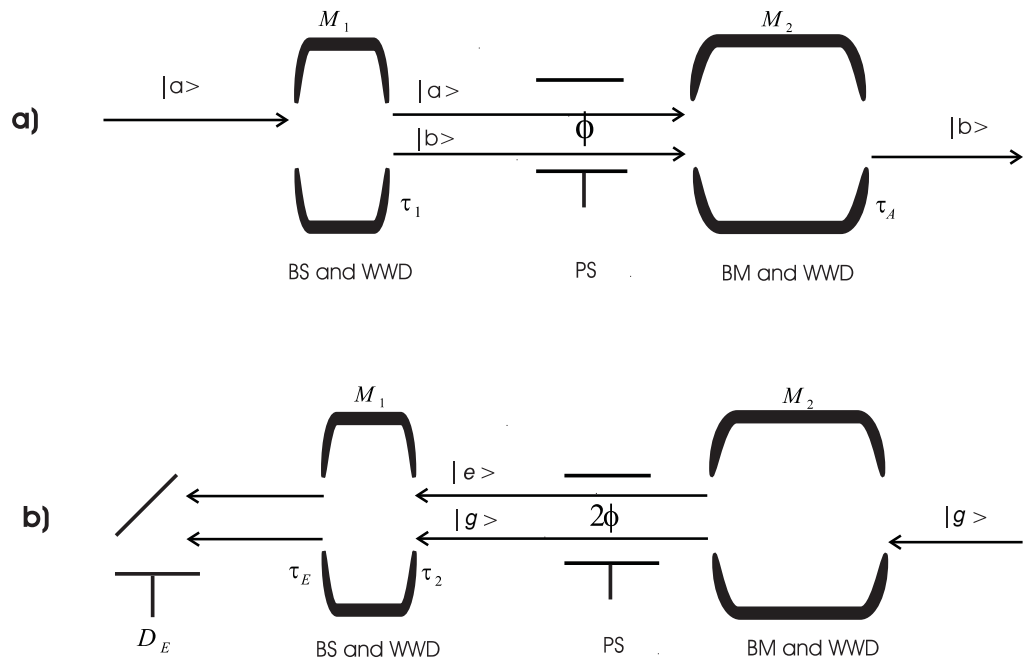}
\end{center}
\caption{\label{fig2} (a) Asymmetric quantum optical Ramsey
interferometer. The Quanton leaves the M$_{1}$ cavity at $t_1$, in
a superposition of two alternatives. The BM is now also quantized,
performing a conditional $\pi $-pulse, leaving the atom with
certainty in state $\mid b\rangle _{A}$ after crossing cavity
M$_{2}$. (b) The atom is launched back to self-erase its own WWI.
At $t_2$ ($t_E$) the Erason enters (leaves) the M$_{1}$ resonator
and is finally measured in $D_{E}$.}
\end{figure}

We follow a different strategy in order to achieve full quantum
erasing which is based in time-reversal invariance. In order to
preserve the time symmetry property of closed systems, the second
microwave pulse providing the beam merger (BM) must be quantized
too. We can use a QORI scheme described in \cite{Davidovich94}. It
consists in two high-Q microwave resonator successively
transversed by a beam of monoenergetic two-level atoms, at such a
low rate that only one atom is present in both cavities at a time
(see Figure \ref{fig2}-a). The symbol ''A'' will be used in the
sequel to label the Quanton. Thus, we denote the upper and lower
levels by $\left\{ |a\rangle _{A},|b\rangle _{A}\right\}$.
Before entering the first cavity the atoms are prepared in the
upper level $\mid a\rangle _{A}.$We choose now an asymmetrical
design for the cavities, as depicted in Figure \ref{fig2}(a). It
consists of two asymmetrical cavities (different lengths), such
that

\begin{eqnarray}
s_{1} &\equiv &\sin (\Omega \tau _{1})\neq 1,  \nonumber \\
s_{2} &\equiv &\sin (\Omega \tau _{2})=1,  \label{2.4}
\end{eqnarray}
where $\tau _{i}$ ( i = 1,2) is the interaction time of the atom
in the cavity M$_{i}.$ The cavities are prepared initially to be
empty, i.e., the state vector of the combined cavities-atom system
is given by

\begin{equation}
\mid \Psi _{0}\rangle =\mid 0,0\rangle \mid a\rangle _{A}
\label{2.5}
\end{equation}

Assuming resonant interaction, the probability from the
upper-to-lower level transition in each cavity is given by the
square of the sine in Eq. (\ref {2.4}). Under this preparation,
the atoms is left with certainty in its lower state after passage
of the second cavity. In this energy transferring situation, there
are alternative routes along which the final lower level state can
be reached. Different routes are realized by changing the photon
number in different cavities. These paths, sketched in Figure
\ref{fig2}(a), will be called the ''upper-'' and ''lower-paths''.
It should be noticed that the conditional dynamics on the
alternative path transitions in M$_{2}$ is due to the mixing of
two different state manifolds of the Jaynes-Cummings interaction.
In the case the atom enters M$_{2}$ in the upper level, it will
undergo a $\pi $-pulse leaving a photon in the cavity for the
upper path. On the other hand, if the atom enter in the lower
level no transitions occur since the atomic-cavity system
coincides with the ground state of the interaction Hamiltonian.

After atom A has exited the cavity M$_{2}$ at time t$_{A}$, the
state of the system reads
\begin{equation}
\mid \Psi (t_{A})\rangle =\{-is_{1}\mid 1,0\rangle +c_{1}e^{-i\phi
}\mid 0,1\rangle \} \otimes \mid b\rangle _{A},  \label{2.6}
\end{equation}
where s$_{1}$ is given in (\ref{2.4}) and
$c_{1}=\sqrt{1-s_{1}^{2}}$. After the passage of a single atom the
cavities have been prepared in a non-local field state, i.e., in a
coherent superposition of correlated states of macroscopically
separated quantum systems. It should be noticed that the stated
preparation give by Eq. (\ref{2.6}) is completely unitary. After
the passage of the pumping atom, the cavity is left, with unit
probability, in a macroscopic superposition with an externally
controllable non-local coherence. Due to the asymmetric design of
the resonators, the phase difference $\phi $ of the interfering
alternatives is mapped, in an unitary fashion, into the non-local
phase of the superposition state of the two WWDs.

Note that we have two WWDs, M$_1$ and M$_2$, both with
$\mathcal{Q}_1=\mathcal{Q}_2=1$. Also, the Predictability in M$_2$
is $\mathcal{P}_2=1$, so the probability of detecting atom A in
the lower level will exhibit a featureless dependence on $\phi .$
However, as will be shown in the next section, fringes of unit
visibility can be extracted from a suitable measurement of the
WWD, provided that full WWI is erased from both cavity fields.

\section{The quantum self-eraser and non-local phase observation.}

We launch backwards the Quanton to erase its own WWI. For
instance, gravity could play the role of an atomic mirror for the
Quanton \cite{Wallis93}. Alternatively, we can take advantage of
the factorization property in (\ref{2.6}) in order to simplify the
experimental setup. As a matter of fact, after t$_{A},$ a
subsequent measure of the Quanton\'{}s internal state can be
performed without introducing projection noise. We can use as the
Erason a replica of the Rydberg two-level atom, with upper and
lower states that we denote by $\left\{ |e\rangle _{E},|g\rangle
_{E}\right\} ,$ from a second beam E (see Figure \ref{fig2}-b).
Thus, the detection of the system-atom in D$_{A}$ can eventually
be used to trigger the excitation laser preparing the Erason in
the lower level $\mid g\rangle _{E}.$ The Erason is sent through
both cavities in reversed order with the same velocity magnitude
as that selected for the Quanton, and thus, it can be regarded as
a velocity reversal version of the Quanton. After calculating the
coherent transients developed by the Erason, the state of the
entangled system is \vspace{0.2cm}
\begin{equation}
\fl\mid \Psi \left( t_{E}\right) \rangle =
\{-i[s_{1}^{2}+c_{1}^{2}e^{-2i\phi }]\mid 0,0\rangle \mid e\rangle
_{E}  +[c_{1}^{{}}s_{1}^{{}}-s_{1}^{{}}c_{1}^{{}}e^{-2i\phi }]\mid
1,0\rangle \mid g\rangle _{E} \}\otimes \mid b\rangle _{A}\ ,
\label{2.7}
\end{equation}
\vspace{0.2cm}
 where t$_{E}$ is the final time at which the Erason
leaves the cavity M$_{1}$. The contributions to the amplitude
coefficients in (\ref{2.7}) can be regarded as labels for the four
possible paths of the global system. Before reaching M$_{1}$, the
paths of the returning Erason perfectly correlate with those of
the Quanton. Thus, a relative phase of 2$\phi $ is accumulated
between the upper and lower atomic paths. At this stage, the role
of the Erason is to probe the which-way information stored in the
cavity M$_{2}$, using time reversal symmetry to transfer it to its
own internal state. However, when the atom crosses cavity M$_{1}$,
a rotation of the Bloch vector is performed such that each of the
two possible outcomes in the subsequent measurement of state of
the Erason is correlated with both upper and lower paths. Due to
the asymmetric design of the interferometer, WWI or quantum
erasure can be obtained by the experimenters by means of letting
the Erason cross one or both cavities, respectively. After
detection of the Quanton, the experimenter may decide at will, in
a delayed-choice fashion, to display either particle or wavelike
aspects of the quantum system, deciding how many cavities the
Erason will be allowed to fly through.

The correlations between the Quanton and the final state of the
Erason after crossing both cavities are given by the final
detection probabilities
\begin{eqnarray}
\mathcal{P}_{ge}^{{}} &=& tr\{\mid e\rangle _{EE}\langle e\mid
\rho (t_{E})\}=c_{1}^{4}+s_{1}^{4}+2c_{1}^{2}s_{1}^{2}\cos (2\phi
), \label{3.2}
\\
\mathcal{P}_{gg}^{{}} &=& tr\{\mid g\rangle _{EE}\langle g\mid
\rho (t_{E})\}=2c_{1}^{2}s_{1}^{2}(1-\cos (2\phi )).
\label{3.2bis}
\end{eqnarray}
After many repetitions of the experiment, the detection
probabilities (\ref{3.2}) will exhibit fringes and antifringes
upon variation of the phase $\phi$. Within repetitions, the
cavities should be reinitialized to the vacuum state. It should
also be noticed that the erasure could be performed at any time
after t$_{A}$ within the lifetime of the cavity, and in particular
after the atom A has been detected \cite{Englert99}. As in the
usual quantum eraser, fringes and antifringes add destructively,
and the featureless pattern
$\mathcal{P}_{ba}=\mathcal{P}_{ge}^{{}}+\mathcal{P}_{gg}^{{}}=1$
is recovered when both Erason detection alternatives are summed
up. On the other hand, in spite the factorization property of the
final state of the Quanton, we would like to remark that the
self-eraser do perform quantum erasing operation. The non-local
phase is constructed on the interfering alternatives of the
Quanton, i.e., its wave-like properties, transferred to the cavity
fields via entanglement swapping. Thus, the full recovery of the
interference fringes in Eqs. (\ref{3.2}, \ref{3.2bis}) demands the
complete erasure of WWI about the particle-like properties of the
quantum stored in the WWDs, as imposed by complementarity.

As discussed in the previous section, the system-atom prepares the
cavity fields in a non-local state. In order to demonstrate the
generation of the macroscopic superposition, interference effects
sensitive to the non-local coherence must be measured, in order to
distinguish this state from an incoherent mixture. The Ramsey
scheme converts the non-local phase into a population difference.
Thus, according to (\ref{3.2}), a simple population measurement
can be used to probe non-local phase of the cavity fields in (
\ref{2.6}). Observation of fringes in the detection probability of
the Erason demonstrates the coherent character of the macroscopic
superposition (\ref {2.6}).

The visibility of the interference pattern implicit in $\mathcal{P}%
_{ge}^{{}} $ of Eqs. (\ref{3.2}) is computed to be
\begin{equation}
{\Large \nu
}_{ge}^{{}}=\frac{2c_{1}^{2}s_{1}^{2}}{c_{1}^{4}+s_{1}^{4}}.
\label{Nu_ge}
\end{equation}

\begin{figure}
\begin{center}
\epsfxsize=26pc 
\epsfbox{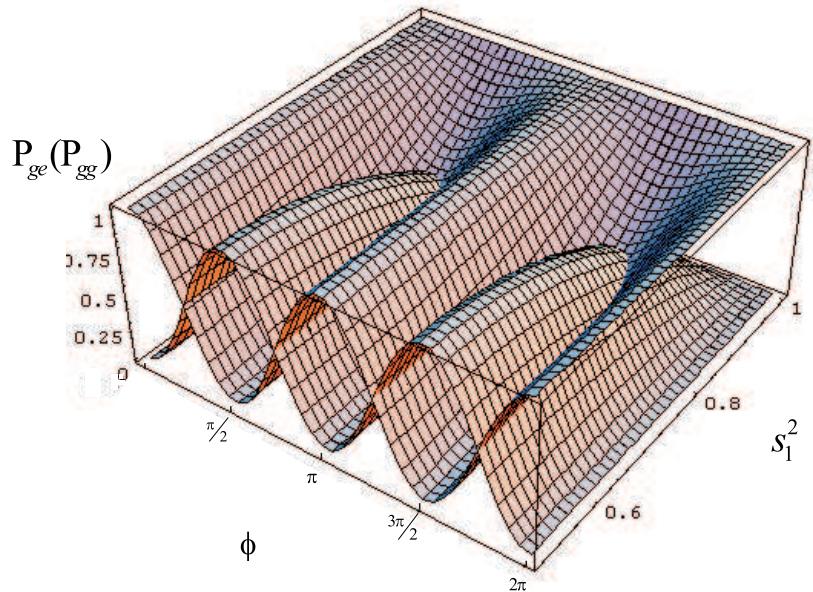}
\end{center}
\caption{\label{fig3}Conditional detection probability
$\mathcal{P}_{ge}^{{}}$ and $\mathcal{P}_{gg}^{{}}$ plotted versus
the externally induce phase $\phi $ and the single cavity
transition probability $s_{1}^{2}$. }
\end{figure}

In the case M$_{1}$ is prepared to act also as a $\pi $-pulse,
$c_{1}s_{1}=0$ and no fringes are obtained as consequence of the
single path situation.
Perfect visibility fringes are obtained for $\mid c_{1}\mid =\mid s_{1}\mid $%
. This situation correspond to a symmetrical mixing
($\mathcal{P}=0$) of the
correlation between eraser and system paths in M$_{1}$. $\mathcal{P}%
_{ge}^{{}}$ and $\mathcal{P}_{gg}^{{}}$, given in Eqs. (\ref{3.2})
are plotted in Figure \ref{fig3} as a function of the induced
phase $\phi $ and of the single-cavity transition probability
s$_{1}^{2}$, showing both visibility limits. Perfect visibility is
obtained for atomic velocities matching the condition

\begin{equation}
v=\frac{4\Omega L_{1}}{\left( 2k+1\right) \pi }\rm{ \qquad \qquad }%
k=0,1,...  \label{3.7}
\end{equation}
where L$_{1}$ is the length of cavity M$_{1}$. Note that Eq.
(\ref{3.7}) is compatible with the $\pi $-pulse preparation of
M$_{2}$, provided this cavity is built twice as long as M$_{1.}$

Further insight into the system can be gained by means of calling
upon time reversal symmetry. The Erason is designed to act as the
velocity reversal version of the Quanton. In fact, time reversal
invariance is used in M$_{2}$ in order to swap the WWI from the
cavity field to the internal state of the atom. However, the
passage coursed by the Erason cannot be regarded as a complete
time reversal of the Quanton passage:\ the open phase shifter
breaks the symmetry, allowing for different alternatives in the
final detection probability. Indeed, the relative phase induced by
the PS, adds up to 2$\phi $ after the Erason passage, instead of
canceling each other. Only in the latter case, the whole
interferometer would exhibit complete time reversal invariance and
the Erason would end up in the initial state of the Quanton. This
can be seen from Eqs. (\ref{3.2}) where, consistently with time
reversal invariance, $\mathcal{P}_{ge}^{{}}=1$ for $\phi $ $=0$.

\section{Violation of temporal Bell inequalities}

A signature of quantum coherence can be found in the violation of
temporal Bell inequalities of the type of \cite{Legget85}. We show
in this section that the quantum self-eraser violates a temporal
Bell inequalities for certain ranges of the Rabi phase. According
to realistic descriptions, dynamical variables possesses definite
values at definite times. Following \cite{Huelga95}, a stochastic
process for a two-state system can be defined as the dichotomic
random variable $\chi (t)$ assuming the values $1(-1)$ when the
system is on upper (lower) state at time $t$. Let us consider now
the quantity\smallskip

\begin{equation}
\,\Delta _{\pm }\rm{=}K_{13}\pm K_{12}\pm K_{23},  \label{Delta}
\end{equation}
where $K_{ij}$ are the two-time autocorrelation function

\begin{equation}
\,K_{ij}\equiv K(t_{i},t_{j})=\left\langle \chi (t_{i})\,\chi
(t_{j})\right\rangle ,  \label{K}
\end{equation}
and we take $t_{1},t_{2}$ as the entering and exit time of the
Quanton through M$_{1\rm{ }}$ ($t_{2}=t_{1}+\tau _{1}$), and
$t_{3}$ as the exit time of the returning Erason after crossing
M$_1$, i.e., $t_{3}=t_{E}$ (see Figure \ref{fig2}). Any two-state
stochastic process must satisfy the inequality \cite{Huelga95}

\begin{equation}
\,-1\le \Delta _{\pm }.  \label{-1Delta}
\end{equation}

\begin{figure}
\begin{center}
\epsfxsize=23pc 
\epsfbox{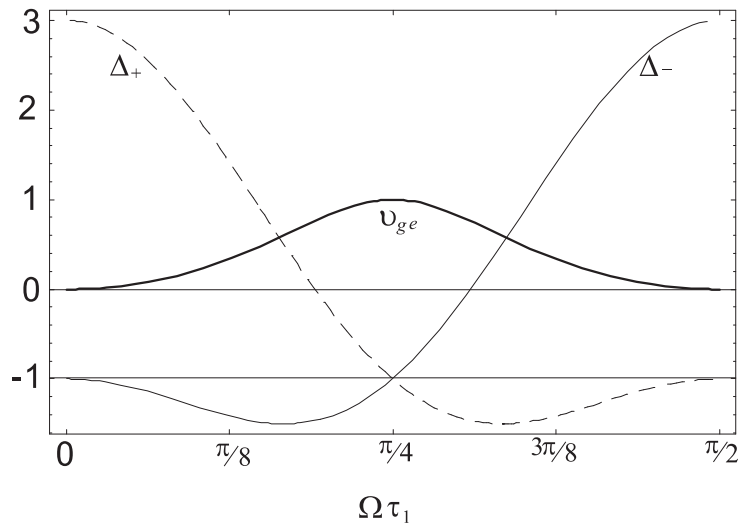}
\end{center}
\caption{\label{fig4} Violation of temporal Bell inequality for
the quantity $\Delta _{\pm }$.}
\end{figure}

On the other hand, from the quantum mechanical point of view, the
autocorrelation function given by $K(t_{i},t_{j})=\left\langle
\sigma _{z}(t_{i})\sigma _{z}(t_{j})\right\rangle $ can be
calculated for the quantum self-eraser to yield

\begin{eqnarray}
K_{12} &=&c_{1}^{2}-s_{1}^{2},  \label{Kaes} \\
K_{23} &=&c_{1}^{4}-s_{1}^{4},  \nonumber \\
K_{13} &=&1-8c_{1}^{2}s_{1}^{2}.  \nonumber
\end{eqnarray}
Inserting (\ref{Kaes}) into (\ref{Delta}), we obtain
\begin{equation}
\,\Delta _{\pm }=\cos (4\Omega \tau _{1})\pm 2\cos (2\Omega \tau
_{1}), \label{Marshall}
\end{equation}
which coincides with the value derived for a single cavity
preparation in \cite{Huelga95}. This coincidence can be understood
taking into account that for any two times $t,t^{\prime }$ between
$t_{2}$ and $t_{3}-\tau _{1}$ we
have $K(t,t^{\prime })=1,$ due to the perfect correlations provided by the $%
\pi $-pulse in M$_{2}.$ The quantity (\ref{Marshall}) is plotted
in Figure \ref{fig4}, showing violation of the inequality
(\ref{-1Delta}) for certain ranges of the vacuum Rabi phase
$\Omega \tau _{1}.$ As can be seen in this plot, there is no
violation at $\Omega \tau _{1}=\pi /2$ (for which $\Delta
_{+}=-1,\Delta _{-}=3),$ which is precisely the point of full
recovery of the interference fringes in the detection probability
of the Erason.

It is also important to remark that Bell experiments involving
Rydberg atoms are especially of interest, since they can close the
communication and the detection loopholes \cite{Loeffler96}. A
proposal to measure Bell's inequality violation with a Rydberg
atom sequentially interacting with two classically driven cavities
has been given in \cite{Kim00}. On the other hand, Rydberg atoms
are mesoscopic systems, which are of interest to study deviations
from macroscopic realism \cite{Huelga96}. Thus, the proposed
self-eraser interferometer posses a wide play-ground to test the
role of measurement in quantum mechanics at the mesoscopic level.

\section{Discussion}

A scheme for a atomic beam quantum self-eraser has been presented.
In contrast to previous models, the present proposal is based on
time reversal invariance. This property demands an experimental
setup in which the Quanton- (Erason-) atoms pump (probe) the
cavities in a symmetrical fashion. The present proposal uses an
asymmetric design of the microwave resonators: the second
resonator allowing one to factorize the state of the exit atom
from the state of the cavities. Thus, we can send another atom
from a reversed beam, to play the role of the self-eraser atom,
provided it is send as a velocity reversal of the pump atom. On
the other hand, the factorizing conditions extend our view about
quantum erasing operation. Interference fringes are recovered
directly: we do not need to correlate the measurements on the
detectors $D_{A}$ and $D_{E}.$ Although in an unconventional
fashion \cite{Kwiat94}, the Erason performs quantum erasing
operation. In fact, due to the experimental configuration, the
actual full recovery of the interference pattern demands the
quantum erasure of WWI from the WWDs, as required by
complementarity. At the point of maximum erasure, we have
$\mathcal{Q}_1=\mathcal{Q}_2=0$, allowing for a full recovery of
the interference pattern ($\mathcal{V}=1$). The self-eraser
highlights the role of complementarity in the recovery of
interference fringes via entanglement swapping.

Also resulting from the asymmetric design of the resonators, the
scheme allows one to prepare the correlated cavity fields in a
macroscopic superposition state in an unitary fashion
\cite{Davidovich94}. When the Quanton leaves the second cavity,
the phase information accumulated by the atom in the interference
region is kept in the cavity fields as the non-local coherence
between the two components of its superposition state. This
coherence is transferred back to the Erason as probed in the
fringe structure of its final state. Observation of an
interference pattern in the final detection probability of the
Erason can be used for a joint demonstration of quantum erasing
operation and the generation of non-local superposition in the
correlated field state of the macroscopically separated cavity
systems. \smallskip

Violation of temporal Bell inequalities provides a criterion for a
signature of the establishment of macroscopic quantum coherence.
On the other hand, the quantum self-eraser is built on
entanglement swapping of macroscopic coherence. Thus, it is
interesting to analyze how both manifestations of quantum
coherence are mutually related. We have shown that the quantum
self-eraser violates temporal Bell inequalities. However, this
violation is not directly connected to quantum erasing capability.
In fact, the point of maximum erasability is outside of the
parameter range leading to violation.

Thus, the quantum self-eraser appears as an useful playground to
study the relation between non-locality, macroscopic realism and
complementarity.

\ack

J. M.-L. was initially supported  by CIC from Universidad
Michoacana de San Nicol\'{a}s de Hidalgo, and the PROMEP\ program
of the Public Education Secretary (SEP) in M\'{e}xico.

\end{document}